\documentclass[10pt,aps]{revtex4}
\usepackage{amssymb}
\usepackage{latexsym}
\usepackage{epsfig}
\usepackage{float}
\usepackage{graphicx,epsfig, color}
\usepackage{graphicx}
\usepackage{subfigure}
\usepackage{amsmath}
\usepackage{epstopdf}
\usepackage{float}
\usepackage{color}
\usepackage[colorlinks,
            linkcolor=blue,
            anchorcolor=blue,
            citecolor=blue]{hyperref}

\begin{document}
\title{\textbf{Thermodynamics of PNJL at zero temperature in a strong magnetic field}}
\author{Yuan Wang, Xin-Jian Wen \footnote{wenxj@sxu.edu.cn} }
\affiliation{ Institute of Theoretical Physics, Shanxi University,
Taiyuan, Shanxi 030006, China} \affiliation{Collaborative Innovation
Center of Extreme Optics, Shanxi University, Taiyuan, Shanxi 030006,
People's Republic of China}

\begin{abstract}
In this paper, the deconfinement and chiral restoration transitions
in strong magnetic field is realized at zero temperature in the
Polyakov Nambu$-$Jona-Lasinio model. We provide the thermodynamic
treatment to mimic the deconfinement phase transition at zero
temperature together with the entangled scalar and vector
interactions coupled with the Polyakov loop. The magnetic catalysis
is found by a rising behavior of the critical chemical potential for
the first-order deconfinement phase transition. While the magnetic
catalysis on the chiral restoration could convert to inverse
magnetic catalysis under the running coupling interaction ansatz.
Furthermore, the stronger magnetic field makes the possible
quarkyonic phase window to be enlarged under the running coupling
interaction.

\end{abstract}



\maketitle

\section{Introduction}
The study of the quantum chromodynamics (QCD) phase diagram attracts
a lot of attention theoretically and experimentally
\cite{Leupold:2011zz,Fukushima:2010bq,DElia:2018xwo,Bali:2011qj}.
The three basic characteristics of QCD, chiral symmetry, quark
confinement and asymptotic freedom, play an important role in
determining the properties of hadrons and the phase diagram at
finite temperature and density. Understanding these aspects could
help us to get a better knowledge of the strongly interacting
matter.

In the past decades, one powerful lattice QCD was established for
2+1 flavors and vanishing baryon chemical potential. It is predicted
that there is a crossover-like transition at high temperature from a
hadronic phase. However, the situation is less clear for finite
chemical potentials due to the well-known difficulty given by the
so-called sign problem, which affects lattice calculations
\cite{Ejiri:2004yw}. Thus the more work on QCD diagram has to be
done in the phenomenological models, which capture the basic physics
of QCD itself and are instructive to evaluate the quark mass, the
pion mass, and so on. First, the asymptotic freedom indicates the
interaction of quarks becomes weaker with decreasing distance and
becomes stronger as the separation increases. It can be described by
a variational coupling constant \cite{Ayala:2018wux} or represented
by an density-and-temperature dependent mass of quasiparticle in
literature \cite{Xia:2018wdj,Wen:2005uf,Xu:2021eyi}. Second, the
chiral symmetry breaking was successfully investigated by the
Nambu-Jona-Lasinio (NJL) model at finite temperature by the
dynamical generation of quark mass, which can act as an order
parameter of chiral phase transition
\cite{Ferrer:2013noa,Costa:2013zca,Nambu:1961tp,Nambu:1961fr,
Buballa:2003qv,Vogl:1991qt,Klevansky:1992qe}. Third, the quark
confinement as an essential feature of QCD involves nonperturbative
properties. The MIT bag model is based on a phenomenological
realization of quark confinement. The bag constant is often
introduced phenomenologically with the expectation that it simulate
non-perturbative corrections
\cite{Chodos:1974je,Chodos:1974pn,DeGrand:1975cf}. At high densities
of a good simulation of the compact star, the deconfinement phase
transition is expected to take place, which is characterized by a
approximately Z(3) center symmetry breaking. On the other hand, the
MIT bag model violates chiral symmetry and the NJL model does not
confine quarks. The question has been addressed by the Polyakov
Nambu-Jona-Lasinio (PNJL) model, where the quarks interact with the
temporal gluon field, represented by the Polyakov loop. The PNJL
model had been successfully employed on the investigation of the
chiral phase diagram and the confinement-deconfinement transition
during a long period \cite{Sasaki:2006ww,Ferreira:2017wtx,
Biswal:2019xju,Carlomagno:2019yvi,Mao:2017tcf}. Unfortunately, the
directly taking the zero temperature limit on the Polyakov potential
in the conventional version of the PNJL model is infeasible, which
will lead to the vanishing of the confinement mechanism. Recently,
by introducing a Polyakov-loop dependent coupling interaction, the
confinement-deconfinement transition in the PNJL model has been
recovered to be operative at the zero temperature regime, which was
named as PNJL0 model \cite{Mattos:2021alf,Mattos:2021tmz}.

The recent investigation of QCD in strong magnetic fields brings a
new sight on the whole phase diagram. The typical strength of the
strong magnetic fields could be of the order of $10^{12}$ Gauss on
the surface of pulsars. Some magnetars can have even larger magnetic
fields as high as $10^{16}$ Gauss at the surface and $10^{18}$ Gauss
in the interior of certain compact stars. By comparing the magnetic
and gravitational energies, the physical upper limit to the total
neutron star is of order $10^{18}$ Gauss \cite{Skokov:2009qp}. And
for the self-bound quark stars, the limit could go higher
\cite{Chanmugam:1992rz,Lai:2000at}. A realistic profile of the
magnetic field distribution inside strongly magnetized neutron stars
is proposed that the magnetic fields increase relatively slowly with
increasing baryon chemical potential in the polynomial form instead
of exponential form \cite{Dexheimer:2016yqu}. At the large hadron
collider energy in CERN, it is estimated to produce a field as large
as $5\times 10^{19}$ Gauss \cite{Kharzeev:2007jp}. Much stronger
background fields might have been produced during the cosmological
electroweak phase transition \cite{Vachaspati:1991nm,Grasso:2000wj}.
The QCD vacuum characterized by the chiral symmetry breaking would
be changed by the enhanced quark-antiquark condensate in strong
magnetic fields, which leads to a dynamical generation of quark
masses. The ccorresponding mechanism is the famous magnetic
catalysis (MC) effect \cite{Gusynin:1994re}. The inverse magnetic
catalysis (IMC) on the (pseudo)critical temperature revealed by the
lattice QCD can be realized by the NJL model with a decreasing
critical temperature as the magnetic field increases
\cite{Ahmad:2016iez,Farias:2014eca,
Ferreira:2014kpa,Farias:2016gmy}. Up to now, the knowledge on QCD
phase diagram is mainly achieved at zero/small chemical potential
and finite temperature. The phenomenological investigations try to
reproduce the lattice result including the (pseudo)critical
temperature and quark condensate at low and high temperature. The
QCD phase in the region of larger chemical potential and zero
temperature has been not well known yet. The aim of the present
paper is to investigate the deconfinement and chiral transition at
zero temperature by improving the thermodynamics treatment of the
PNJL0 model in strong magnetic fields. Of special interest is the
effect of the magnetic field on the critical chemical potential, the
confinement property dependent on the Polyakov loop at zero
temperature.

This paper is organized as follows. In Section \ref{sec:model}, we
present the thermodynamics of the magnetized quark matter in SU(2)
PNJL0 model. In Section \ref{sec:result}, the numerical results for
the chiral symmetry restoration and deconfinement phase transition
are shown at zero temperature. The discussions are focused on the
magnetic effect on the chiral and the deconfinement transition with
the magnetic field independent and dependent coupling constants. The
last section is a short summary.

\section{General formalism of PNJL0 in strong magnetic fields}\label{sec:model}

Following the work in the SU(2) version of the PNJL model, the
Lagrangian density in a strong magnetic field is given by
\cite{Mattos:2021alf,Hansen:2006ee}
\begin{equation}
{\mathcal{L}}_{\mathrm{PNJL}}=\bar{\psi}(i\gamma_\mu D^\mu -m)\psi
+G_s[(\bar{\psi}\psi )^{2}-(\bar{\psi}i\gamma _{5}\vec{\tau}\psi
)^{2}]-G_v(\bar
{\psi}\gamma_{\mu}\psi)^2 -U(\Phi,\bar{\Phi},T).
\end{equation}%
where $\psi $ represents a flavor isodoublet ($u$ and $d$ quarks) and $%
\vec{\tau}$ are isospin Pauli matrices. The coupling of the quarks
to the electromagnetic field is introduced by the covariant
derivative $D_\mu=\partial_\mu-ieQA_\mu$, where
$Q=\mathrm{diag}(q_u,q_d)=\mathrm{diag}(2/3,-1/3)$ is the quark
electric charge matrix. The Polyakov potential describes the
deconfinement at finite temperature. In literature, the effective
potential $U(\Phi, \bar{\Phi}, T)$ exhibits a phase transition from
color confinement to color deconfinement.

At finite temperature, the Polyakov potential depends explicitly on
the traced Polyakov loop and its conjugate $\Phi$ and $\bar{\Phi}$
\cite{Roessner:2006xn,Dutra:2013lya}. In order to obtain the
confinement description at zero temperature, we take
$\Phi=\bar{\Phi}$ for the nonzero quark chemical potentials at the
mean-field-approximation. So the total thermodynamical potential
density for the two-flavor quark matter in the mean-field
approximation reads
\begin{eqnarray}
\Omega_{\mathrm{PNJL}}&=&\sum_{i=u,d}\Omega_{i}
+G_s\sigma^2-G_v\rho^2+U(\Phi,T),\label{omega}
\end{eqnarray}
The integral in the effective thermodynamics potential is not
convergent. In literature, the Fock-Schwinger proper-time method
\cite{Schwinger:1951nm} is applied to a thermal field theory to
obtain the exact expression of the effective potential with
B-dependent divergent part in the vacuum regularization
\cite{Ebert:1999ht,Ayala:2016bbi,Abramchuk:2019lso}. Another
equivalent method is the magnetic field independent vacuum
regularization (MFIR), which is widely employed in the recent work
\cite{Menezes:2008qt,Menezes:2009uc,Avancini:2011zz,Allen:2013lda,Farias:2015eea,Rabhi:2011mj,Chatterjee:2011ry}.
In this paper, we adopt the second regularization scheme. The first
term $\Omega_i$ is defined as $\Omega _{i}=\Omega
_{i}^{\mathrm{vac}} +\Omega _{i}^{\mathrm{mag}} +\Omega
_{i}^{\mathrm{med}}$. The vacuum, the magnetic field, and medium
contributions to the thermodynamical potential are
\cite{Menezes:2008qt, Avancini:2011zz, Menezes:2009uc}
\begin{eqnarray}
\Omega_i ^{\mathrm{vac}}&=&\frac{N_{c}}{8\pi ^{2}}\left[ M^{4}\ln
(\frac{\Lambda +\epsilon _{\Lambda }}{M})-\epsilon _{\Lambda
}\Lambda (\Lambda ^{2}+\epsilon _{\Lambda }^{2})\right] ,\\
\Omega_i ^{\mathrm{mag}}&=&-\frac{N_{c}|q_ieB|^2}{2\pi ^{2}}\left[
\zeta'(-1,x_i) -\frac{1}{2}(x_i^2-x_i)\ln(x_i)+\frac{x_i^2}{4}\right] ,\\
\Omega_i ^{\mathrm{med}}&=&-\frac{T |q_i|
eB}{2\pi^2}\sum_{n_i=0}^\infty \alpha_{n_i}\int_{0}^{\infty }(
z_\Phi^+ + z_\Phi^- ) dp.\label{eq:omemed}
\end{eqnarray}
where the quantity $\epsilon _{\Lambda }$ is defined as $\epsilon
_{\Lambda}=\sqrt{\Lambda ^{2}+M^{2}}$ and  $x_i=\frac{M_{i}^2}{2
|q_i| B}$ is dimensionless. The spin degeneracy factor
${\alpha}_{n}=2-\delta_{n0}$ is 1 for the lowest Landau level (LLL)
and 2 for otherwise
higher Landau levels. The ultraviolet divergence in the vacuum part $%
\Omega_i ^{\mathrm{vac}}$ of the thermodynamical potential is
removed by the momentum cutoff. The partition function densities
$z_\Phi^\pm$ are evaluated by the color traces \cite{Hansen:2006ee}
\begin{eqnarray}z_\Phi^+&=&\ln\{1+3(\bar{\Phi}+\Phi e^{-\frac{E_i+\tilde{\mu}_i}{T}})e^{-\frac{E_i+\tilde{\mu}_i}{T}} +
e^{-3\frac{E_i+\tilde{\mu}_i}{T}}\} , \\
z_\Phi^-&=&\ln\{1+3(\Phi+\bar{\Phi}
e^{-\frac{E_i-\tilde{\mu}_i}{T}})e^{-\frac{E_i-\tilde{\mu}_i}{T}} +
e^{-3\frac{E_i-\tilde{\mu}_i}{T}}\}.
\end{eqnarray}
Only $z_\phi^-$ would survive at zero temperature, which produces a
traditional step function in the medium term as \cite{Menezes:2008qt} \cite{Menezes:2009uc}
\begin{eqnarray}\Omega_i^{med}&=&-\frac{|q_i| eB}{2\pi^2}\int_0^{p_z^F}
3(\tilde{\mu}_i-E_{ni}) \notag\\
&=& -\frac{N_c |q_i|eB }{4\pi^2} \sum_{n_i=0}^{n^{\mathrm{max}}_{i}}
{\alpha}_{n_i} \left\{ \tilde{\mu}_i
\sqrt{\tilde{\mu}_i^2-M^2_{ni}}-M^2_{ni}\ln[\frac{\tilde{\mu}_i+\sqrt{\tilde{\mu}_i^2-M^2_{ni}}}{M_{ni}}]
\right\}\label{eq:ome_med},
\end{eqnarray}
where $M_{ni}=\sqrt{M_i^2+2 n |q_i eB|}$ and the color degenerate
factor 3 is recovered once more due to the decouple of the color
interaction with the polyakov potential.

By minimizing the thermodynamical potential with respect to the
quark condensate $\sigma_i$ and the Polyakov loop $\Phi$, we can
have a set of the coupled gap equations
\cite{Buballa:2003qv,Mattos:2021alf}
\begin{eqnarray} \frac{\partial P}{\partial \sigma}=0, \ \ \frac{\partial P}{\partial\tilde{\mu}}=0, \ \
\frac{\partial P}{\partial \Phi}=0. \label{eq:equation}
\end{eqnarray}
Therefore, we can have an equivalent system of non-interacting quark
with the constitute dynamical mass $M$ and renormalized chemical
potential $\tilde{\mu}$ \cite{Buballa:2003qv}.
\begin{eqnarray}
M_i&=&m_{i0}-2G_s\sigma_i.  \label{eq:gap}\\
\tilde{\mu}_{i}&=&\mu_i-2G_v\rho_{i}.
\end{eqnarray}
In our work, the isospin symmetry is assumed and we have
$M_u=M_d=M$. At zero temperature, the occupied Landau levels have
the maximum value
\begin{equation}
n_{i}^{max}=\frac{{ \tilde{\mu}_i}^2-M^2}{2|q_i| eB}.
\label{eq:landau}
\end{equation}

The second term in the Eq. (\ref{omega}) is the contribution from
the quark condensate $\sigma=\sum_{i=u,d}\sigma_i$. The condensation
contribution from the quark with flavor $i$ is
\begin{equation}
\sigma_{i}=\sigma_{i}^{\mathrm{vac}}+\sigma_{i}^{%
\mathrm{mag}} +\sigma_{i}^{%
\mathrm{med}}.  \label{eq:condensate}
\end{equation}
The terms $\sigma_i ^{\mathrm{vac}}$, $\sigma_i ^{\mathrm{mag}}$ and $\sigma_i ^{%
\mathrm{med}}$ represent the vacuum, the magnetic field, and medium
contributions to the quark condensation, respectively as following
\cite{Avancini:2011zz, Menezes:2009uc},
\begin{eqnarray}
\sigma_i ^{\mathrm{vac}} &=&-\frac{MN_{c}}{2\pi ^{2}}\left[ \Lambda \sqrt{%
\Lambda ^{2}+M^{2}}-M^{2}\ln (\frac{\Lambda +\sqrt{\Lambda ^{2}+M^{2}}}{M})%
\right], \\ \label{eq:convac}
\sigma _{i}^{\mathrm{mag}} &=&-\frac{M|q_{i}|eBN_{c}}{2\pi
^{2}}\left\{ \ln [\Gamma (x_{i})]-\frac{1}{2}\ln (2\pi)+x_{i}
-\frac{1}{2}(2x_{i}-1)\ln
(x_{i})\right\} , \\
\sigma _{i}^{\mathrm{med}} &=& \frac{M|q_{i}|eBN_{c}}{%
2\pi ^{2}} \sum_{n_{i}=0}{\alpha}_{n_{i}}\int_0^\infty \frac{dp_z}{E_{ni}}\Theta(\tilde{\mu}_i-E_{ni})\notag \\
&=&\frac{M|q_{i}|eBN_{c}}{%
2\pi ^{2}} \sum_{n_i=0}^{n^\mathrm{max}_{i}} {\alpha}_{n_i} \ln
\Big(\frac{\tilde{\mu}_i+\sqrt{\tilde{\mu}_i^2-M^2_{ni}}}{M_{ni}}\Big).
\end{eqnarray}

The simple polynomial form for the Polyakov potential was improved
by replacing the higher order polynomial term with the logarithm
form \cite{Fukushima:2003fw,Roessner:2006xn,Roessner:2007gha}. At
finite temperature, the following ansatz is suggested
\cite{Dexheimer:2008av, Dexheimer:2009hi}
\begin{eqnarray}{\cal{U}}(\mu,\Phi)=(a_0 T^4+a_1 \mu^4+a_2 T^2 \mu^2)\Phi^2+a_3
T_0^4 \ln(1-6\Phi^2+8\Phi^3-3\Phi^4) .
\end{eqnarray}
At zero temperature, we adopt the following formula
\begin{eqnarray}
{\cal{U}}_0(\mu,\Phi)\equiv a_1\mu^4\Phi^2+a_3 T_0^4 \ln(1-6
\Phi^2+8\Phi^3 -3\Phi^4),
\end{eqnarray}
where $T_0=190$ MeV is very often used as the critical temperature
for deconfinement in the PNJL model \cite{Ratti:2005jh}. At zero
temperature, the ${\cal{U}}_0$ is importantly to the existence of
confinement-deconfinement transition.

In literature, the four-quark vertex of one-gluon exchange diagram
was changed into the entangled vertex \cite{Kondo:2010ts}. Inspired
by this phenomenology, the PNJL model was extended by introducing an
entangled interaction between the quark condensate and the traced
Polyakov loop in EPNJL model \cite{Sakai:2010rp}, where the chiral
resotration and the deconfinement transition is produced
simultaneously in agreement with the Lattice result.  Recently, the
entanglement interaction dependent on the traced Polyakov loop in
both the Polyakov potential and the effective interaction between
quarks was introduced to avoid the lack of confinement physics in
PNJL at $T=0$ \cite{Mattos:2021alf}. In strong magnetic fields, we
employ the phenomenology by making the scalar and vector interaction
dependent on the traced Polyakov loop as
\begin{eqnarray}
G_s \rightarrow G_s(1-\Phi^2), \ \  G_v \rightarrow
G_v(1-\Phi^2).\label{eq.coupl}
\end{eqnarray}
Therefore, the effective coupling interaction would vanish in the
deconfined phase due to the dependence relation in Eq.
(\ref{eq.coupl}).

At $T=0$, the pressure is given by
\begin{eqnarray}P=-{\cal{U}}(\sigma,\rho,\mu,\Phi)+G_v\rho^2-G_s
\sigma^2-\sum_{i=u,d}\Omega_{i}
\end{eqnarray}
with the Polyakov potential
\begin{eqnarray}
{\cal{U}}(\sigma,\rho,\mu,\Phi)&=&{\cal{U}}_{0}(\mu,\Phi)-G_s{\Phi}^2
{\sigma}^2+ G_v {\Phi}^2 {\rho}^2 .
\end{eqnarray}

From the thermodynamics potential one can easily obtain the quark
number density $\rho=\sum_{i=u,d}\rho_i$ with the $i$ flavor
contribution
\begin{eqnarray}\rho_i=\frac{3 |q_i|eB}{2 \pi^2}\sum_{n_{i}=0}{\alpha}_{n_{i}}\sqrt{\tilde{\mu}_i^2-M_i^2-2 n_i |q_i|
eB} .
\end{eqnarray}

\section{Numerical result and conclusion}\label{sec:result}
The important prediction of the QCD is the thermodynamic transition
at sufficient high temperature and/or high density from the hadron
phase to the color-deconfined quark-gluon plasma. The chiral
transition and deconfinement transition is depicted by the
well-defined order parameters. The quark condensate and the
non-vanished Polyakov loop value are solved by the coupled gap
equations as well as to minimize the thermodynamics potential. In
this section our investigation of the QCD thermodynamics is
restricted to the zero temperature. The four-fermion coupling
interaction in the model is adopted with the fixed constant and
magnetic field dependent running coupling in the following
subsections respectively.

\subsection{Results with fixed coupling constant}
Being the non-renormalizable model, a regularization procedure is
usually applied by a three-momentum non-covariant cutoff
$\Lambda=587.9$ MeV.  The quark current mass as free parameters are
adopted as $m_u=m_d=5.6$ MeV. The four-fermion couplings are
$G_s=2.44/\Lambda^2$, and $G_v=0.3G_s$. We adopt the parameters as
$a_1=-0.05$ and $a_3=-0.2$ for the confinement potential guided by
the Ref. \cite{Mattos:2021alf}. The presence of the vector
interaction was discussed for the realization of the deconfinement
transition at zero magnetic field. In this section, we would discuss
the effect of the magnetic field on the deconfinement and chiral
transition.

\begin{figure}[H]
\centering
\includegraphics[scale=0.35]{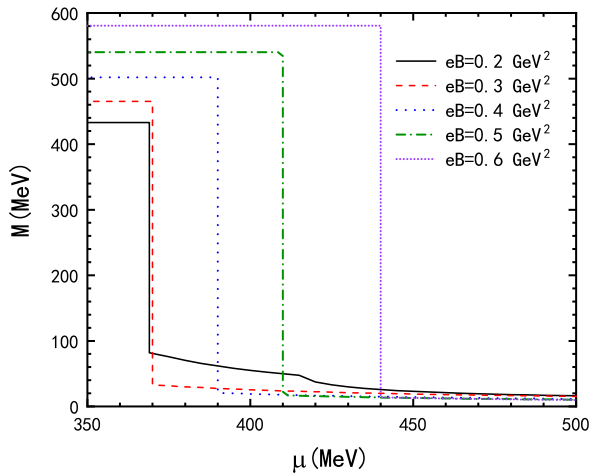}
\caption{\small{The dynamical mass as a function of the chemical
potential $\mu$ for several magnetic fields $eB$ in unit of
GeV$^2$.} \label{fig:fig1}}
\end{figure}

\begin{figure}[H]
\centering
\includegraphics[width=0.4\textwidth]{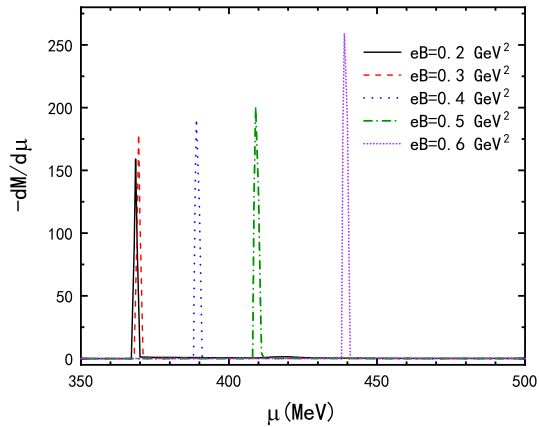}
\caption{\label{fig:fig2}The susceptibility $-dM/d\mu$  as a
function of chemical potential $\mu$ , for several magnetic fields
$eB$ in unit of GeV$^2$. }
\end{figure}

The dynamical mass is widely accepted as the order parameter of the
chiral transition, which can be solved from the gap equation. In
Fig. \ref{fig:fig1}, the dynamical mass $M$ for $u$- and $d$-quarks
is shown as a function of the chemical potential $\mu$. The
increases of the magnetic field is marked by the lines from the
bottom $0.2$ GeV$^2$ to the top $0.6$ GeV$^2$. It is clearly that in
the chiral broken phase of small chemical potentials, the larger
magnetic fields would result in larger dynamical masses. This
catalyzing effect of magnetic field on the dynamical chiral symmetry
breaking is known as the so-called MC effect
\cite{Miransky:2015ava}. In all the lines there is a sudden falling
behavior, representing the appearance of the first-order phase
transition from the chiral broken phase to the restoration. It
should be emphasized that the chiral phase transitions are always
the first-order at zero temperature for various magnetic fields. The
explicit chemical potential for the chiral restoration can be showed
by the peak of the derivative of the $M$ with respect to the
chemical potential $\mu$. In Fig. \ref{fig:fig2}, the susceptibility
$-\frac{dM}{d \mu}$ is shown as functions of the chemical potential.
It is found that the peak of $-\frac{dM}{d \mu}$ moves to the larger
chemical potentials as the magnetic fields increase. It is again
emphasized that the magnetic field has a strong tendency to enhace
the quark-antiquark condensates, namely reflecting the MC effect at
zero temperature.

\begin{figure}[H]
\centering
\includegraphics[width=0.4\textwidth]{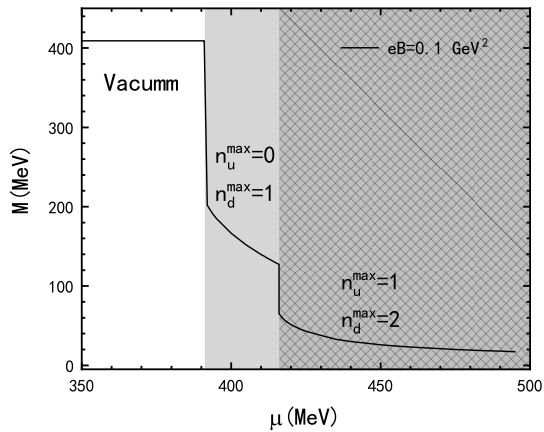}
\caption{\label{fig:fig3}The dynamical mass and the maximum Landau
level as a function of chemical potential $\mu$ at the magnetic
field $eB=0.1$ GeV$^2$. }
\end{figure}

In the above Figs.\ref{fig:fig1} and \ref{fig:fig2}, the magnetic
field is sufficiently large to suppress all the quarks to the LLL.
The main contribution to the quark condensate should come from the
quarks at the LLL. For the lowest nonzero value of $|eB|$, the zero
Landau level occupation is not significant. The sum over more Landau
level has to be taken. In Fig. \ref{fig:fig3}, the dynamical mass is
calculated at the $eB=0.1$GeV$^2$. The two first-order transitions
are resulted by the presence of a mismatch in the maximum Landau
level for $u$ and $d$ quarks given by the $n_u^\mathrm{max}$ and
$n_d^\mathrm{max}$. It is clear that as the density increases, there
are more Landau levels occupied by quarks but the $n_d^\mathrm{max}$
is higher than $n_u^\mathrm{max}$ in the weak magnetic field. It
would have an influence on the trend of the magnetic catalysis on
the critical chemical potential, which will be given in later
section.

\begin{figure}[H]
\centering
\includegraphics[width=0.4\textwidth]{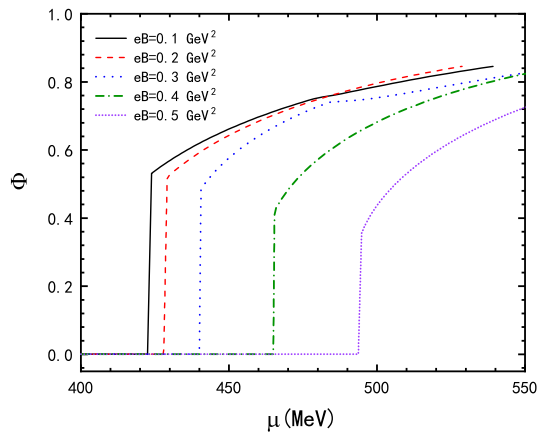}
\caption{\label{fig:fig4}The Polyakov loop $\Phi$ as a function of
chemical potential $\mu$, for several magnetic fields
$eB$ in unit of GeV$^2$. }
\end{figure}

In Fig. \ref{fig:fig4}, the Polyakov loop $\Phi$ as an order
parameter of the deconfinement transition is shown versus the
chemical potential at different magnetic fields. It is clearly seen
that for all magnetic fields the values of $\Phi$ have a sudden jump
to around 0.5 from zero and then continue to increase as the
magnetic field increases. This emphasized that there are first-order
phase transitions from the confinement to the deconfined phases at
zero temperature. A noteworthy point is the non-monotonic hehavior
and the intersect with each other of the $\Phi$ in the deconfined
phase at low magnetic fields $eB=0.1$, 0.2, and $0.3$ GeV$^2$, which
is caused by the Landau energy level. The Landau energy level is
dominated by the values of dynamical mass $M$, chemical potential
$\mu$, and the magnetic field. The value of the Landau energy level
would more sensitively depend on the change of $(M^2-\mu^2)$ at the
weak magnetic field.

\begin{figure}[H]
\centering
\includegraphics[width=0.4\textwidth]{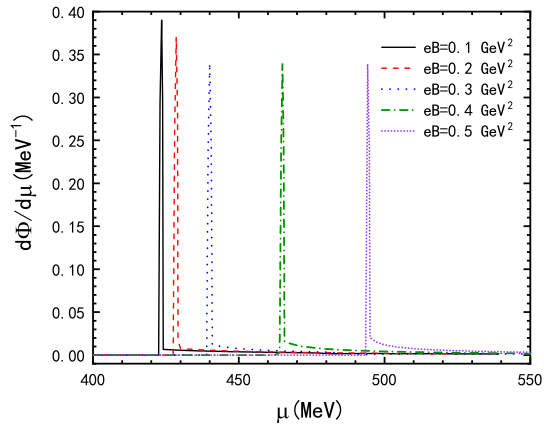}
\caption{\label{fig:fig5}The derivative of $\Phi$ with respect to
$\mu$ as a function of chemical potential $\mu$, for several
magnetic fields $eB$ in unit of GeV$^2$. }
\end{figure}

Correspondingly, the derivative of $\Phi$ with respect to the
chemical potential $\mu$ as a function of chemical potential $\mu$
are shows in Fig. \ref{fig:fig5}. The critical chemical potential
$\mu_c^\Phi$ can be exactly signaled by the peak of $d\Phi/d\mu$. It
is worth noting that the deconfinement transition at zero
temperature is produced in strong magnetic field in our model.
Moreover, the first-order phase transition occurs in region of
larger $\mu_c^\Phi$ at more stronger magnetic field. This also
implies a MC effect for the deconfinement transition.

\subsection{Results with magnetic-field-dependent coupling constant}

Furthermore we discuss the results under the
magnetic-field-dependent coupling constant. The coupling becomes
weak at stronger magnetic fields due to the asymptotic freedom
mechanism \cite{Coleman:1973sx}. We chose the magnetic
field-dependent coupling constant ansatz in Ref.
\cite{Ferreira:2017wtx}\cite{Ferreira:2014kpa}:
\begin{equation}
G_s(eB)=G_s^0 {\frac{1+a(\frac{eB}{\Lambda^{2}_\mathrm{QCD}})^2+
b(\frac{eB}{\Lambda^{2}_\mathrm{QCD}})^3}{1+c(\frac{eB}{\Lambda^{2}_\mathrm{QCD}})^2
+d(\frac{eB}{\Lambda^{2}_\mathrm{QCD}})^4}}.
\label{eq:running}
\end{equation}%
In our work, the parameters are fixed as:
$\Lambda_\mathrm{QCD}=300\mathrm{MeV}, a=0.0108805,
b=-1.0133\times10^{-4}, c=0.02228, d=1.84558\times10^{-4} $
\cite{Ferreira:2017wtx}\cite{Ferreira:2014kpa}.

\begin{figure}[H]
\centering
\includegraphics[width=0.4\textwidth]{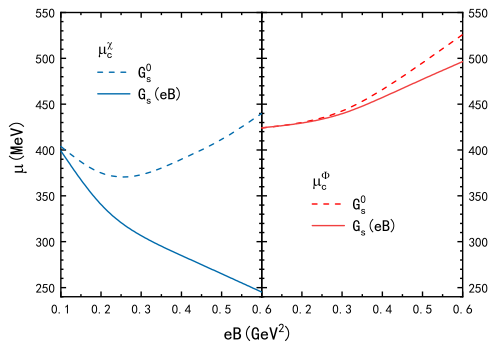}
\caption{\label{fig:fig6}The critical chemical potentials of chiral
(left panel) and deconfinement phase transitions (right panel) as
the function of $eB$, with  a constant coupling $G_s^0$ and a
magnetic field dependent coupling $G_s(eB)$ from Eq.
(\ref{eq:running}). }
\end{figure}
The critical chemical potentials for the chiral and deconfinement
phase transitions are plotted as the functions of $eB$ with two
kinds of couplings in Fig. \ref{fig:fig6}. The blue lines on the
left panel show the results $\mu_c^\chi$ of the chiral phase
transitions. According to the statement in the preceding section,
there are two first-order transitions at $eB=0.1$ GeV$^2$. The
average of two chemical potentials is taken as the critical chemical
potential that plotted in Fig. \ref{fig:fig6}. It is obviouis that
at the fixed coupling $G_s$ the critical chemical potential marked
by the blue-dashed line is decreased firstly and then goes up as the
magnetic field increases. There is a clear qualitative similarity
with the results in Refs.
\cite{Ferreira:2017wtx,Miransky:2015ava,Ferreira:2015jrm}. In the
region of sufficiently small magnetic fields and larger coupling
constant, the critical chemical potential decreases with the field
in Fig. 6 from Ref. \cite{Miransky:2015ava}. And in the weak
magnetic field regime, the critical chemical potential would show a
temporary decrease as the magnetic field increases at $T=0$ in Fig.
7.4 of Ref. \cite{Ferreira:2015jrm}. However, as the magnetic field
goes up to much larger value, the magnetic catalysis effect on the
chiral chemical potential becomes obvious. By employing the running
coupling interaction $G_s(eB)$, the $\mu_c^\chi$ marked by the
blue-solid line would go down drastically. This shows a visible
difference between the two kinds of interactions. The trend of
$\mu_c^\chi$ with $eB$ indicates the so-called IMC effect, which was
realized by the behavior of the decreasing critical temperature as
the $eB$ increases \cite{Preis:2010cq}. The magnitude of decrease of
$\mu_c^\chi$ is almost qualitatively in agreement with the results
in Ref. \cite{Ferreira:2017wtx} at zero temperature. On the right
panel, the critical chemical potential $\mu_c^\Phi$ depicted by the
red lines is steadily growing as the magnetic field increases under
the both couplings $G_s^0$ and $G_s(eB)$. It can be seen that the
magnetic field dependence of the coupling $G_s(eB)$ in Eq.
(\ref{eq:running}) would have weak influence on the deconfinement
transition. It illustrates that the deconfinement transition depends
insensitively on the running behavior of the coupling constant. The
trend of $\mu_c^\Phi$ always keeps a monotonical increasing
function. On the contrary, it is also observed that for the chiral
transition, the difference in $\mu_c^\chi$ between the couplings
$G_s(eB)$ and $G_s$ becomes larger as the magnetic field increases,
which indicates that the chiral restoration is more sensitive to the
strength of the applied magnetic field through the scalar
interaction.
\begin{figure}[H]
\centering
\includegraphics[width=0.4\textwidth]{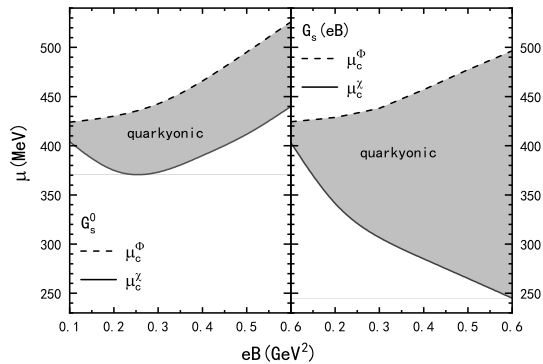}
\caption{\label{fig:fig7}The critical chemical potentials
$\mu_c^\chi$, $\mu_c^\Phi$, and the possible quarkyonic phase window
at zero temperature in strong magnetic fiels, with a constant
coupling $G_s^0$ (left panel) and a magnetic field dependent
coupling $G_s(eB)$ (right panel). }
\end{figure}
In literature, the quarkyonic phase was proposed as a new phase of
QCD and expected to exist at large chemical potentials in the chiral
symmetry restoration but confined region
\cite{Mao:2018wqo,Fukushima:2008wg,Abuki:2008nm,McLerran:2008ua,Buisseret:2011ms},
which indicates that the chiral restoration transition occurs
earlier than the deconfinement transition as the chemical potential
increases. To facilitate this observation, the critical chemical
potential for the chiral phase transition and the deconfinement
phase transition are compared in Fig. \ref{fig:fig7}. Results with a
fixed coupling constant $G_s$ are shown on left panel and with a
magnetic-field-dependent coupling constant $G_s(eB)$ on right
panel. The critical chemical potential $\mu_c^\chi$ with both
couplings $G_s$ and $G_s(eB)$ are always lower than the
$\mu_c^\Phi$ for the deconfinement phase transition. Consequently,
there is a region $\mu_c^{\mathcal{X}} < \mu< \mu_c^\Phi$ is
identified as the quarkyonic phase, where the chiral symmetry is
partial restored while quarks are still confined. It can be seen
that the quarkyonic phase window marked by the gray region is
enlarged as the magbnetic field increases. Moreover, by comparison
with the fixed coupling $G_s$, the magnetic-field-dependent coupling
constant $G_s(eB)$ on the right panel would lead to a larger
expansion of the quarkyonic window at stronger magnetic fields.

\section{Summary}
In this paper, we have investigated the deconfinement phase
transition and chiral phase transition at zero temperature by
improving the PNJL0 model to the strong magnetic field. The
interaction of quarks is described by the Polyakov potential
together with the entanglement of the scalar and vector
interactions. The chiral restoration and the deconfinement
transitions take place as the chemical potential increases. The
strong magnetic field would play an important role in the phase
transition. It has been found that the critical chemical potential
for the deconfinement moves to the larger value as the magnetic
field increase, which indicate the MC effect no matter how the
coupling constant is employed. Moreover, the critical chemical
potential $\mu_c^\Phi$ becomes more insensitive to the magnetic
field under the running coupling $G(eB)$. In contrary, the effect of
the magnetic field on the chiral restoration would sensitively
depend on the coupling. At larger magnetic fields, it has been shown
that the MC effect with the magnetic-field independent coupling
$G_s$ would convert to the IMC with the running coupling $G_s(eB)$.
Finally, it has been verified that the quarkyonic phase window is
present under the condition $\mu_c^\chi<\mu_c^{\Phi}$, where the
chiral symmetry is restored but the quarks are still confined.
Moreover, the quarkyonic phase window is evidently enlarged by the
running coupling $G_s(eB)$ at stronger magnetic fields. Since the
previous phenomenological model mainly concentrate on the knowledge
of QCD at finite temperature, we expect that our work at zero
temperature can be regarded as a complementary to the QCD diagram,
where the critical chemical potential would be important to the
dense star matter.

\acknowledgments{ The authors would like to thank support from the
National Natural Science Foundation of China under the Grant Nos.
11875181, 11705163 and 12147215. This work was also sponsored by the
Fund for Shanxi "1331 Project" Key Subjects Construction.}

\end{document}